\documentclass[aps,prd,preprint,amsmath,superscriptaddress]{revtex4}

\usepackage{graphicx}

\begin{document}
\title{ Detecting the dark matter annihilation at the ground EAS detectors}

\author{X.-J. Bi}
\email{bixj@mail.ihep.ac.cn}
\affiliation{Key laboratory of particle astrophysics, IHEP, 
Chinese Academy of Sciences, Beijing 100049, P. R. China}
\author{Y.-Q. Guo}
\affiliation{Key laboratory of particle astrophysics, IHEP,
Chinese Academy of Sciences, Beijing 100049, P. R. China}
\author{H.-B. Hu}
\affiliation{Key laboratory of particle astrophysics, IHEP,
Chinese Academy of Sciences, Beijing 100049, P. R. China}
\author{X. Zhang}
\affiliation{Theoretical Physics Division, IHEP, 
Chinese Academy of Sciences, Beijing 100049, P. R. China}

\begin{abstract}
In this paper we study the possibility of detecting gamma rays
from dark matter annihilation in the subhalos of the Milky Way
by the ground based EAS detectors within the frame of the minimal
supersymmetric standard model. Based on the Monte Carlo simulation 
we also study the properties of two specific EAS detectors, 
the ARGO and HAWC, and the sensitivities
of these detectors on the detection of dark matter annihilation.
We find the ground EAS detectors have the possibility to observe such signals.
Conversely if no signal observed we give the constraints on the supersymmetric
parameter space, which however depends on the subhalos properties.
\end{abstract}

\maketitle

\section{ introduction }

The existence of cosmological dark matter has been established by
a multitude of observations.  
The evidences come mainly from the gravitational effects of the dark
matter component, such as the observations of the rotation curves in
spiral galaxies and velocity dispersion in elliptical galaxies, the
X-ray emission and peculiar velocities of galaxies in the clusters of
galaxies, all indicating much steeper gravitational potentials than those
inferred from the luminous matter. The primordial nucleosynthesis and
cosmic microwave background measurements constrain the baryonic component
to be about 4\% of the critical density, while the total amount of the clumpy
matter is about 30\% of the critical density. 
Therefore most of the dark matter is of non-baryonic origin. 

The nature of the non-baryonic dark matter is one of the most outstanding
puzzles in particle physics and cosmology.
However, the gravitational effects do not shed light on the solutions of
this problem. Eventhough, some hints can still be obtained on the nature
of dark matter.
The simulation of structure formation requires the existence of dark matter
and favors the nature of cold dark matter (CDM), that is, the dark matter
particles are non-relativistic when they freeze out the thermal bath
at the early universe. The CDM nature rules out the candidate of neutrino
as the dominant component of dark matter since neutrinos are hot (relativistic)
when they freeze out at the temperature of about $1 MeV$. 
The precise measurement of the abundance of
the dark matter component also constrains the nature of dark matter
by requiring a natural explanation of the measured density. The three years
WMAP data, combining with recent observational results from other
experiments, give the abundance of CDM as $\Omega_{\text{CDM}}h^2
 = 0.109^{+0.003}_{-0.006}$ \cite{wmap}. The small error strongly constrains
the dark matter models.

All the candidates of non-baryonic dark matter
require physics beyond the standard model of particle physics. 
Among the large amount of candidates, the most attractive
scenario involves the weakly interacting massive particles (WIMPs).
An appealing idea is that the WIMPs form the thermal relics of the
early universe and naturally give rise to the relic abundance in the
range of the observed values for both the interaction strength and
the masses being at the weak scale. The WIMPs are also well motived
theoretically by the physics beyond the standard model to solve the
hierarchical problem between the weak scale and the Planck scale.
In particular, the minimal supersymmetric extension of the standard model
(MSSM) provides an excellent WIMP candidate as the lightest supersymmetric 
particle, usually the neutralino, which are stable due to R-parity 
conservation \cite{jungman}. The cosmological constraints on the 
supersymmetric (SUSY) parameter space have been extensively studied in
the literature \cite{csusy}.

Another appealing aspect of WIMP is that it can be detected 
on the present running or proposed experiments, either directly
by measuring the recoil energy when WIMP scatters off the
detector nuclei \cite{direct} or indirectly by observing the
annihilation products of the WIMPs, 
such as the antiprotons, positrons, $\gamma$-rays or 
neutrinos \cite{indirect}. The WIMPs may also be generated in
the next generation colliders, which is the most direct way to 
resolve the nature of the dark matter particles. The direct and indirect
detection of dark matter particles are viable and complementary
ways to the collider studies in order to further constrain 
the nature of dark matter.
WIMP annihilation provides viable explanation for exotic signals
observed in the cosmic ray experiments, such as the `GeV excess'
of the Galactic diffuse $\gamma$ observed by EGRET \cite{egret}, 
the bump at about 10 GeV of the positron ratio
measured by HEAT \cite{heat} and the TeV $\gamma$-ray emission 
at the Galactic
center observed by HESS \cite{hess} and CANGAROO II \cite{cangaroo}.

The rate of the WIMP annihilation is proportional to the number
density square of the dark matter particles. Therefore the 
searches for the annihilation signals should aim at the 
regions with high matter densities,
such as at the galactic center \cite{gc} or the nearby subhalos
\cite{subhalo,kou}. The existence of a wealth of subhalos throughout the
galaxy halos is a generic prediction of the CDM paradigm of structure
formation in the Universe. High resolution simulations show that
for CDM scenario the large scale structure forms hierarchically by continuous
merging of smaller halos and as the remnants of the merging process
about 10\% to 15\% of the total mass of the halo is in the form
of subhalos \cite{tormen98,klypin99,moore99,
ghigna00,springel01,zentner03,delucia04,kravtsov04}.
At the center of the subhalos there are high mass densities
and therefore they provide alternative sites for the search of WIMP 
annihilation products.

There are several advantages
of detecting the $\gamma$-rays from the subhalos than that from the GC.
First, subhalos produce clean annihilation signals, while the 
annihilation radiation from the 
GC is heavily contaminated by the baryonic processes associated with
the central supermassive black hole (SMBH) and the supernova remnant
Sgr A$^*$ \cite{Zaharijas}. 
Furthermore, the dark matter density profile near the GC is complicated due
to the existence of baryonic matter and leads to difficulties in making
theoretical calculations. For example, the SMBH can
either steepen or flatten the slope of the DM profile at the innermost
center of the halo depending on the evolution of the black hole
\cite{ullio}. For subhalos, their profile may
simply follow the simulation results. Second, the small
subhalos form earlier and have larger concentration parameter, which
leads to relatively greater annihilation fluxes.
Third, the DM profile may be not universal, as shown in the simulation
given in Ref. \cite{reed,jing}. Smaller subhalos have steeper central cusp.
In this case, if taking the GC the NFW profile and the
subhalos the Moore profile, the $\gamma$-ray fluxes from the
subhalos may even be greater than that from the GC.
Forth, according to the hierarchical formation of structures
in the CDM scenario we expect that subhalos should contain their
own smaller sub-subhalos, which can further enhance the annihilation
flux. The sub-subhalos have been observed in the numerical simulations,
such as in the Ref. \cite{zentner05}.
Finally, the environmental trend seems to make the subhalos more concentrated
\cite{bullock01}. However, the effects need further studies by more precise
simulations.

The possibility of detecting dark matter annihilation signal from the 
Galactic Center (GC) has been extensively studied \cite{gc}. The 
high energy $\gamma$-rays from the GC observed by HESS and CANGAROO II have
been explained as a possible signal of dark matter annihilation 
\cite{hpf}. The subhalos are approximately uniformly distributed 
in the Milky Way dark halos and provide potential $\gamma$-ray sources which
will be observed in the next generation experiments. However, the position
of the subhalos can not be predicted by numerical simulations,
therefore the search for the $\gamma$-rays from the subhalos 
need detectors with large field of view. Unless we have known the position
of a nearby subhalo the \v{C}erenkov detectors can not do the job of
blind search due to their narrow field of view ($\sim 5^\circ$),
despite they have high sensitivities. The satellite based experiments,
such as the GLAST \cite{glast} or the AMS \cite{ams}, usually have large
field of view. The possibility to detect
$\gamma$-rays from the subhalos by GLAST has been studied \cite{kou,peirani}.
However, the satellite based instruments have small effective area 
at the order of $\sim 1 m^2$, which limits their ability to detect
low $\gamma$ ray fluxes. Therefore the ground based experiments,
which can have very large effective area, are complementary to the
satellite based experiments.

In this paper, we explore the possibility of detecting $\gamma$-ray
signals from subhalos by the ground based EAS detectors, such as the
ARGO \cite{argo} and the next generation All-Sky VHE Gamma-Ray 
Telescope HAWC \cite{hawc}. 
These detectors have large field of view and large effective area.
In the case that the neutralino is heavy, which annihilates
into $\gamma$-rays with high energy and low intensity, the ground
based EAS detectors can
be even superior than the satellite based experiments.
 
In the next section we first describe our model for the subhalo distribution
and the particle nature of dark matter. The fluxes of gamma rays from the
subhalos are then calculated. In the Sec III, we discuss how the
ground EAS detectors can constrain the SUSY models and the properties
of two EAS detectors and the sensitivity of dark matter detection.
We finally give summary and conclusion in the Sec IV.

\section{ model description }

The flux of gamma rays from the neutralino annihilation in the subhalos
is given by
\begin{equation}
\label{flux}
\Phi(E)=\phi(E)\frac{\langle\sigma v\rangle}{2m^2}
\int{dV \frac{\rho^2}{4\pi d^2}} = \frac{\phi(E)}{4\pi}
\frac{\langle\sigma v\rangle}{2m^2}\times \frac{1}{d^2} 
\int_{0}^{\bar{r}}4\pi r^2 \rho^2(r)dr\ ,
\end{equation}
where $\phi(E)$ is the differential flux at energy $E$
by a single annihilation in unit of $1$ particle $GeV^{-1}$,
$m$ is the mass of the dark matter particle,
$d$ is the distance from the detector to the source,
$\bar{r}=\text{min}(R_{sub}, r_{\Delta \Omega})$ is the
minimal value of the subhalo radius $R_{sub}$ and the 
angular radius at the distance $d$ within the angular
resolution $\Delta \Omega$ of the detector.
We notice that the integration in Eq. (\ref{flux}) depends
only on the distribution of the dark matter $\rho(r)$, taken as a
spherically-averaged form,  which is determined
by numerical simulation or by observations and has no relation
to the particle nature of the dark matter. We define this factor as
`cosmological factor' and the first part in Eq. (\ref{flux})
the `particle factor' which is exclusively determined by its particle
nature, such as the mass, strength of interaction and so on.
                                                                                
The cosmological factor in Eq. (\ref{flux}) is determined by
the position, mass and interior profile  of the subhalo.
We adopt the N-body simulation results to calculate the
cosmological factor.

\subsection{distribution of the subhalos}

N-body simulation is widely adopted
to investigate the spatial distribution and mass function of
substructures in the host halo. The results show that the radial
distribution of substructures is generally
shallower than the density profile of the smooth background
due to the tidal disruption of substructures
which is most effective near the galactic center \cite{diemand}.
The relative number
density of subhalos can be approximately given by an isothermal
profile with a core \cite{diemand}
\begin{equation}
\label{dis}
n(r)=2n_H(1+(r/r_H)^2)^{-1}\ ,
\end{equation}
where $n_H$ is the relative number density at the scale
radius $r_H$, with $r_H$ being about $0.14$ times the halo virial radius
$r_H=0.14r_{\text{vir}}$. 
The result given above agrees well with that in another recent
simulation by Gao et al. \cite{gao}.

Simulations show that the differential mass
function of substructures has an approximate power law
distribution, $dn/dm\sim m^{-\alpha}$.
In Ref. \cite{diemand} both the cluster and galaxy substructure
cumulative mass functions are found to be an $m^{-1}$ power law,
$n_{\text{sub}}(m_{\text{sub}} > m)\propto m^{-1}$,
with no dependence on the mass of the parent halo.
A slight difference is found in a recent simulation by Gao \textit{et al.}
\cite{gao} that the cluster substructure is more abundant than galaxy
substructure since the cluster forms later and more substructures
have survived the tidal disruption.
The mass function for both scales are well fitted by $dn/dm\propto m^{-1.9}$.
Taking the power index of the differential mass function
smaller than -2 makes the fraction of the total mass enclosed in
subhalos insensitive to the mass of the minimal subhalo we take.
The mass fraction of subhalos estimated in the literature is around
between 5 percent to 20 percent \cite{ghigna00,springel01,stoehr}.
In this work we will always take the differential index of $-1.9$ and the
mass fraction of substructures as 10 percent.

We then get the number density of a
substructure with mass $m$ at the position $r$ to the galactic center
\begin{equation}
\label{prob}
n(m,r)=n_0 \left(\frac{m}{M_{\text{vir}}}\right)^{-1.9} (1+(r/r_H)^2)^{-1}\ ,
\end{equation}
where $M_{\text{vir}}$ is the virial mass of the MW, $n_0$ is the
normalization factor determined by requiring the total mass of
substructures converges to 10 percent of the virial mass of the MW.
A population of substructures within the virial radius of the
MW are then realized statistically due to the probability of Eq. (\ref{prob}).
The mass of the substructures are taken randomly between
$M_{\text{min}}=10^6 M_\odot$, which is the lowest substructure
mass the present simulations can resolve \cite{cut}, and the maximal mass
$M_{\text{max}}$.
The maximal mass of substructures is taken to be $0.01M_{\text{vir}}$
since the MW halo does not show recent mergers of satellites with masses
larger then $\sim 2\times 10^{10}M_\odot$.
The $\gamma$-ray flux is quite insensitive to
the minimum subhalo mass since the flux from a single subhalo
scales with its mass \cite{kou,aloisio}.

However, due to the finite resolution of the N-body simulations
the distribution in Eq. (\ref{prob}) is an extrapolation of the
subhalo distribution at large radius. The formula
underestimates of the tidal effect which destroys most substructures
near the GC. We take the tidal effects into
account under the ``tidal approximation'', which assumes that
all mass beyond the tidal radius is lost in a single
orbit while keep its density profile inside the tidal radius intact.
                                                                                
The tidal radius of the substructure is defined as the radius 
at which the tidal forces of the host exceeds the self gravity of the
substructure. Assuming that both the host and the substructure gravitational
potential are given by point masses and considering the centrifugal
force experienced by the substructure the tidal radius at the Jacobi limit
is given by \cite{hayashi}
\begin{equation}
r_{\text{tid}} = r_c \left( \frac{m}{3M(<r_c)} \right)^{\frac{1}{3}}\ ,
\end{equation}
where $r_c$ is the distance of the substructure to the GC,
$M(<r_c)$ refers to the mass within $r_c$.

The substructures with $r_{\text{tid}} \lesssim r_s$ will be disrupted.
The mass of a substructure is also recalculated by subtracting the mass
beyond the tidal radius.
After taking the tidal effects into account we find 
the substructures near the GC are disrupted completely.  The
substructures with NFW profile can exist more near the GC than the Moore
profile. This is because that the NFW profile has smaller $r_s$.

\subsection{ concentration parameter }

We will adopt both the NFW and Moore profiles of dark matter 
distribution in our study. The NFW profile was first prosed by
Navarro, Frenk, and White \cite{nfw97} and supported by
recent studies\cite{nfws} that the DM profile of isolated and relaxed halos
can be described by a universal form
\begin{equation}
\label{nfw}
\rho_{DM}(r)= \frac{\rho_s}{(r/r_s)(1+r/r_s)^2}\ ,
\end{equation}
where $\rho_s$ and $r_s$ are the scale density and scale radius respectively.
The two free parameters of the profile can be determined by
the measurements of the virial mass of the halo
and the concentration parameter determined by simulations.
The concentration parameter is defined as
\begin{equation}
c=\frac{r_{vir}}{r_{-2}}\ ,
\end{equation}
where $r_{vir}$ is  the virial radius of the halo and $r_{-2}$ is the radius
at which the effective logarithmic slope of the
profile is $-2$, i.e., $\frac{d}{dr}(r^2\rho(r))\left|_{r=r_{-2}}=0\right . $.
For the NFW profile we have $r_s=r_{-2}$.
The concentration parameter reflects how the DM
is concentrated at the center.

Moore \textit{et al.} gave another form  of the DM
profile \cite{moore} to fit their numerical simulation
\begin{equation}
\label{moore}
\rho_{DM}(r)= \frac{\rho_s}{(r/r_s)^{1.5}(1+(r/r_s)^{1.5})}\ ,
\end{equation}
which has the same behavior at large radius as the NFW profile while
it has a steeper central cusps $\rho(r) \to r^{-1.5}$ for small $r$ than
the NFW profile.
The index of the central cusp at about $1.5$ is also favored by following
higher resolution simulations\cite{moores}.
For the Moore profile we have $r_s=r_{-2}/0.63$.
                                                                                
\begin{figure}
\includegraphics[scale=0.9]{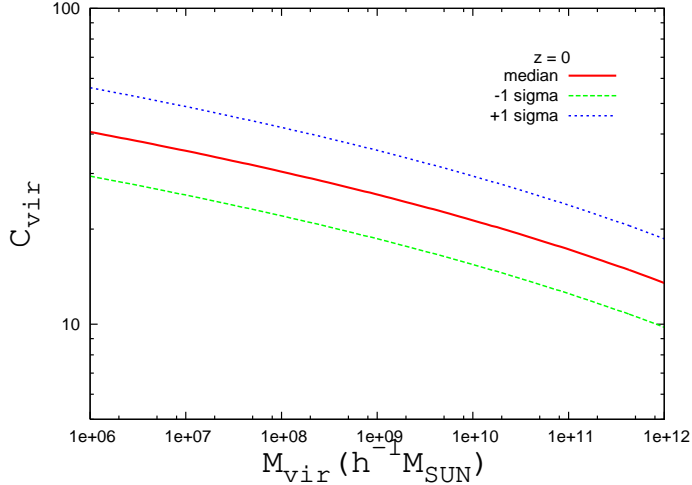}
\caption{\label{concen}
Concentration parameter as a function of the virial mass
calculated according to the Bullock model\cite{bullock01}.
The model parameters are taken as $F=0.015$ and $K=4.4$.
The cosmology parameters are taken as  $\Omega_M=0.3$, $\Omega_\Lambda=0.7$,
$\Omega_Bh^2=0.02$, $h=0.7$, $\sigma_8=0.9$ with three generations of
massless neutrinos and a standard scale invariant 
primordial spectrum.  Both the median and the $\pm 1\sigma$ values of 
the concentration parameters are plotted.
}
\end{figure}

Concentration parameter is obtained by simulation. In a semi-analytic model
based on their simulation results Bullock et al.\cite{bullock01} 
found that the concentration of a halo
is strongly correlated with the formation epoch of the halo.
At an epoch of redshift $z_c$ a typical collapsing mass $M_{*}(z_c)$
is defined by $\sigma[M_{*}(z)]=\delta_{sc}(z)$, where the
 $\sigma[M_{*}(z)]$ is the linear rms density fluctuation on the
comoving scale encompassing a mass $M_*$, $\delta_{sc}$ is the
critical overdensity for collapsing  at the spherical collapse model.
The model assumes the typical collapsing mass is related to
a fixed fraction of the virial mass of a halo $M_{*}(z_c)=FM_{\text{vir}}$.
The concentration parameter of a halo with virial mass $M_{\text{vir}}$
at redshift $z$ is then determined as
$c_{\text{vir}}(M_{\text{vir}},z)=K\frac{1+z_c}{1+z}$. Both $F$ and $K$
are constants to fit the numerical simulations.
A smaller $M_{\text{vir}}$
corresponds to a smaller collapsing mass and early collapsing epoch when
the Universe is denser and therefore a larger concentration parameter.
Fig. \ref{concen}
plots the concentration parameter at $z=0$ as a function of the
virial mass of a halo according to the Bullock model \cite{bullock01}.

\begin{figure}
\includegraphics[scale=0.75]{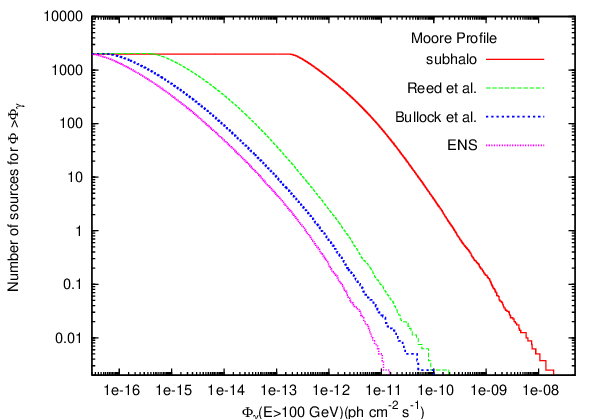}
\includegraphics[scale=0.75]{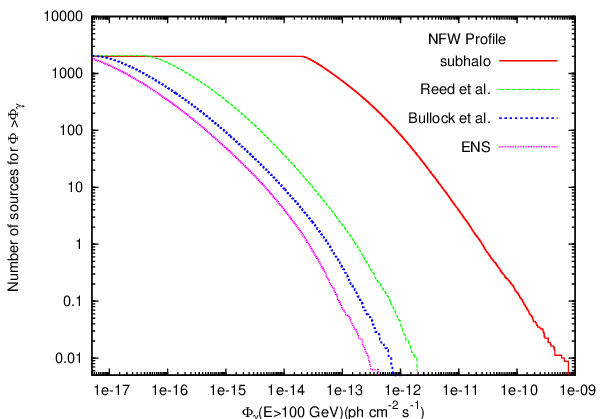}
\caption{\label{result}
The cumulative number of subhalos as function of the
integrated $\gamma$-ray fluxes $n(>\Phi_\gamma)$ for the Moore
profile (left panel) and the NFW profile (right panel).
The results are given within the zenith angle of $60^\circ$.
The curves represent the results according to
different simulations as explained in the text.
These curves give the number of subhalos which emit $\gamma$-rays
with the integrated flux above $\Phi_\gamma$.
}
\end{figure}

From Fig. \ref{concen} we can see that between the masses
$10^6 M_\odot\sim 10^{10} M_\odot$ an experiential
formula $c_{\text{vir}}\propto M_{\text{vir}}^{-\beta}$ 
reflect the simulation result accurately. We expect that this exponential
relation should
be very well followed, since small halos form early at the epoch
when the Universe is dominated by matter with approximate
power-law power spectrum of fluctuations\cite{bullock01}.
However, when we fit the formula to other recent simulation 
results in the literature we find quite large difference,
especially for subhalos from distinct small halos.
We adopt these results to calculate
the  density profile of the substructure and furthermore the
$\gamma$-ray flux from the substructure, which have large uncertainties,
see Fig. \ref{result}.
We find the concentration parameter is the most sensitive parameter
in determining the $\gamma$-ray flux.

By realizing one hundred MW sized halos distributed with subhalos
due to Eq. \ref{prob} we calculate the average
gamma ray intensities from the MW subhalos \cite{bi}.
Fig. \ref{result} gives the cumulative number of subhalos emitting
$\gamma$-rays with the integrated flux above 100 GeV greater than
a value $\Phi_\gamma$. In the calculation we take the particle factor
fixed so that the gamma ray flux from the Galactic Center is just
below the experimental limit, i.e., $\phi_{\gamma}=10^{-9} cm^{-2}s^{-1}$
for the Moore profile.
In the left
panel we plot the results for the Moore profile while the right
panel is for the NFW profile. The curves are given by calculating
the density profile of subhalos according to different
author's N-body simulation results, where
`subhalo' denotes the simulation results
given in Ref. \cite{bullock01} for real subhalos in 
a large smooth dark halo (the dense matter environment);
`Reed et al.' refers to the simulation results given by Reed et al 
\cite{reed}; 
`Bullock et al.' uses the median $c_{vir}-M_{vir}$ relation for 
distinct halos of the Bullock model given in Ref. \cite{bullock01};
`ENS' refers to the result by Eke, Navarro and Steinmetz \cite{eke} for the
$\Lambda$CDM model with $\sigma_8=0.9$. The latter three models actually
do not describe real subhalos. Instead they describe the distinct halos
with small masses. Therefore we expect the following studies refer to
`subhalos' may give somewhat more realistic results.
From Fig. \ref{result}  we can easily read the number
of the expected detectable subhalos if the sensitivity of a
detector is given with same threshold energy and angular resolution
adopted here.

\subsection{SUSY parameter}

We now turn to the particle factor in Eq. (\ref{flux}).
We will work in the frame of MSSM, the low energy
effective description of the fundamental theory at the
electroweak scale. 
By doing a random scan we find how the parameter space is constrained 
by the ground EAS detectors. 

For the R-parity conservative MSSM,
the lightest supersymmetric particle (LSP), generally the
lightest neutralino, is stable and an ideal candidate of dark matter.

However, there are more than one hundred free SUSY breaking
parameters even for the R-parity conservative MSSM.
A general practice is to assume some relations between the parameters
and greatly reduce the number of free parameters.
For the processes related with dark matter production and annihilation,
only seven parameters are relevant under some simplifying assumptions,
i.e., the higgsino mass parameter $\mu$, the wino mass parameter $M_2$,
the mass of the CP-odd Higgs boson $m_A$, the ratio of the Higgs
Vacuum expectation values $\tan\beta$, the scalar fermion mass parameter
$m_{\tilde{f}}$, the trilinear soft breaking parameter $A_t$
and $A_b$. To determine the low energy spectrum of the SUSY particles
and coupling vertices,
the following assumptions have been made: all the sfermions
have common soft-breaking mass parameters $m_{\tilde{f}}$; all trilinear
parameters are zero except those of the third family; the bino and wino
have the
mass relation, $M_1=5/3\tan^2\theta_W M_2$, coming from the unification
of the gaugino mass at the grand unification scale.
                                                                                
We perform a numerical random scanning of the 7-dimensional
supersymmetric parameter space using the package DarkSUSY
\cite{darksusy}. The ranges of the parameters are as following:
$50 GeV < |\mu|,\ M_2,\ M_A,\ m_{\tilde{f}} < 10 TeV$,
$1.1 < \tan\beta < 55$, $-3m_{\tilde{q}} < A_t, \ A_b < 3m_{\tilde{q}}$,
$\text{sign}(\mu)=\pm 1$.
The parameter space is constrained by the theoretical consistency
requirement, such as the correct vacuum breaking pattern,
the neutralino
being the LSP and so on. The accelerator data
constrains the parameter further
from the spectrum requirement, the invisible Z-boson width,
the branching ratio of $b\to s\gamma$ and the muon magnetic moment.

The constraint from cosmology is also taken into account
by requiring the relic abundance of neutralino
$0 < \Omega_\chi h^2< 0.124$,
where the upper limit corresponds to the 5$\sigma$ upper bound from
the cosmological observations. When the relic abundance of neutralino
is smaller than a minimal value we can assume two different cases.
One is that the neutralino relic is produced thermally and
represents a subdominant dark matter component.
In this case we rescale the galaxy neutralino density
as $\rho(r)\to\xi\rho(r)$ with
$\xi=\Omega_\chi h^2/(\Omega_\chi h^2)_{\text{min}}$. We take
$(\Omega_\chi h^2)_{\text{min}}=0.079$, the 5$\sigma$ lower bound
of the CDM abundance \cite{wmap}.
The effect of coannihilation between the fermions is taken into account
when calculating the relic density numerically.
The other case is assuming the neutralino relic is determined
by a nonthermal mechanism \cite{nonthermal}. 
In this case the dark matter is all made up by neutralino.
                                                                                
The  $\gamma$-rays from the neutralino annihilation arise mainly
in the decay of the neutral pions produced in the fragmentation processes
initiated by tree level final states, the quarks, leptons and gauge
bosons. The fragmentation and decay processes
are simulated with Pythia package\cite{pythia}
incorporated in DarkSUSY. We focus our calculation
on the continuum $\gamma$-rays from the pion decays.
                                                                                
Fig. \ref{result} shows the gamma ray emission from subhalos due to 
a fixed particle factor. In the present work we scan the SUSY 
parameter space and study the variation of $\gamma$-ray
flux from the subhalos. We then explore how 
the SUSY parameter space is constrained by the ground EAS detectors
when observing the $\gamma$-ray emission from subhalos.

\section{ observation of neutralino annihilation from subhalos }

In this section, we study the observation of $\gamma$-rays from 
neutralino annihilation in subhalos by the ground based EAS detectors. 
We first show how the SUSY models can be constrained by the 
EAS detectors in an experiment-independent way by
assuming its sensitivity.
Then we will discuss two specific examples of such detectors, the
ARGO and HAWC experiments, and their ability on the dark matter detection.

At present there are two kinds of different techniques adopted 
by the ground-based gamma ray detectors:
the air \v{C}herenkov telescopes (ACTs) and the extensive air
shower (EAS) detectors. There has been great progresses
in improving the sensitivity of the ACTs in the recent years.
However, they have narrow
field of view and can only view a small region of the sky
at any one time. The ACTs can only operate on clear moonless nights
and constrain their observation efficiency.
On the other hand, the EAS detectors, such as the Tibet Array 
\cite{amenomori1999} and the
Milagro observatory \cite{milagro}, can view the entire overhead sky and
operate continuously. To search the unknown $\gamma$-ray sources, such as
the unknown AGNs or subhalos of the MW, EAS detectors with improved
sensitivities are appropriate instruments.

The detectability of a signal is defined by the ratio of
the signal events to the fluctuation of the background.
Since the background follows the Poisson statistics, its
fluctuation has the amplitude proportional to $\sqrt{N_B}$.
The \textit{significance} of the detection is quantified by
$\sigma=\frac{n_\gamma}{\sqrt{N_B}}$.
                                                                                
The signal events are given by
\begin{equation}
\label{ngamma}
n_\gamma= \epsilon_{\Delta \Omega}\int_{E_{th},\Delta \Omega}^{m_\chi} A_{eff}(E,\theta)
\phi(E) dE d\Omega dT\ \ ,
\end{equation}
where $\epsilon_{\Delta \Omega}=0.68$ is the fraction of signal
events within the angular resolution of the instrument and the
integration is for the energies above the threshold energy $E_{th}$ to
the cutoff of the spectrum at the neutralino mass, within
the angular resolution of the instrument $\Delta \Omega$ and for the
observation time. 
The $\phi(E)$ is the flux of $\gamma$-rays from DM annihilation.
The effective
area $A_{eff}$ is a function of energy and zenith angle.

The corresponding
expression for the background is similar to Eq. (\ref{ngamma}) by
substituting the signal spectrum with the background spectrum and
also the effective area to that of cosmic ray background.
The background includes contributions from the hadronic and electronic
comic-rays and the Galactic and extragalactic $\gamma$-ray emissions,
which are given in \cite{bi}. Since the nuclei background dominates
other contributions we only consider the nuclei background in this
work.

\subsection{constrain the SUSY model by EAS detectors -- a general discussion}
\label{gdis}

\begin{figure}
\includegraphics[scale=1.4]{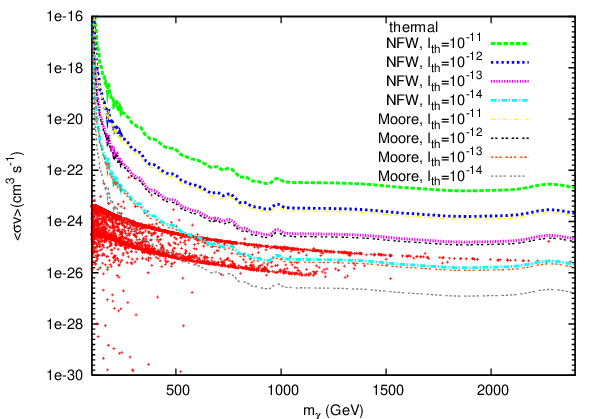}
\includegraphics[scale=1.4]{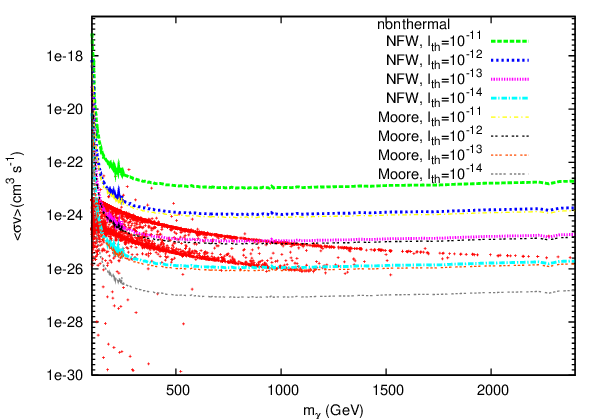}
\caption{\label{general}
Constraints on the SUSY parameter space 
by EAS detectors with different sensitivities $I_{th}$.
The upper panel gives the constraints
assuming that the dark matter are produced thermally while
the lower is for nonthermal production.
Both NFW and Moore profiles are adopted for the subhalos.
}
\end{figure}

\begin{figure}
\includegraphics[scale=1.4]{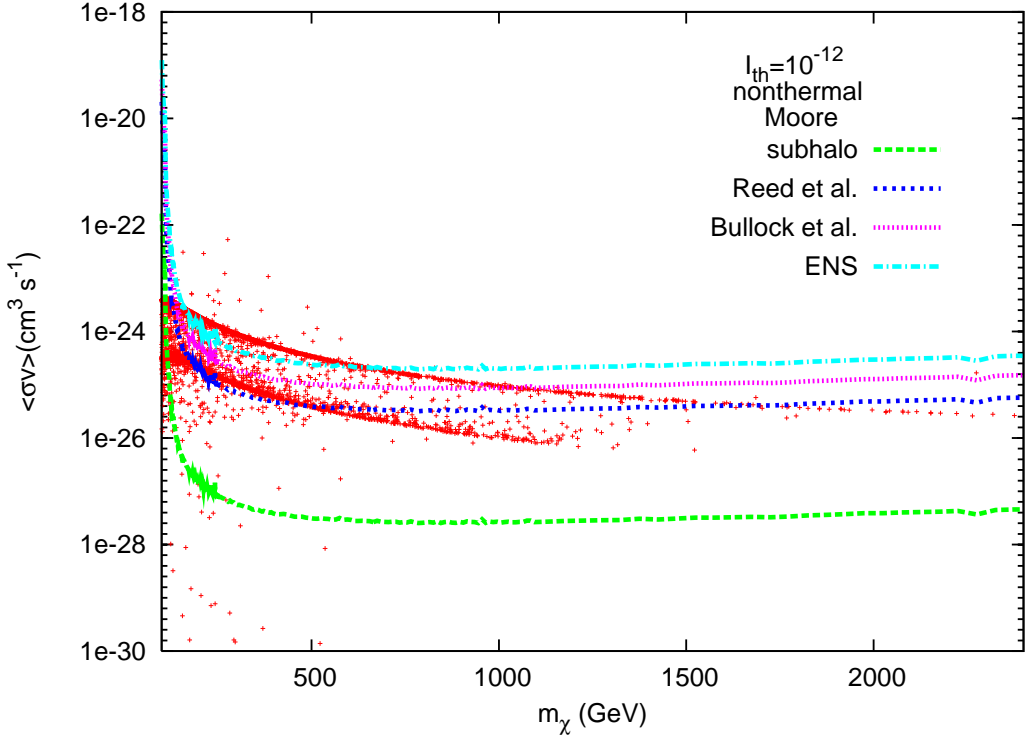}
\caption{\label{gene}
Constraints on the SUSY parameter space 
by EAS detectors with sensitivity $I_{th}=10^{12}$ 
photons $cm^{-2} s^{-1}$ for different subhalo models.
Nonthermal production and Moore profile are adopted for the subhalos.
}
\end{figure}

In this subsection we try to give a general discussion of how
the EAS detectors can constrain the SUSY parameter space.
According to Fig. \ref{result} we have known the cumulative numbers
of the $\gamma$-ray sources generated by dark matter annihilation
in subhalos for different subhalo models. The result is given 
for a specific SUSY model. Once the sensitivity of an EAS detector 
is known we can predict that how many $\gamma$-ray sources are expected
to be detected by the detector and how the number of $\gamma$-ray sources
varies with SUSY model, thanks to the fact, see Eq. (\ref{flux}), that
the `particle factor' and the `cosmological factor' are separated. 
Conversely if no such gamma sources are
found we can constrain the strength of the $\gamma$-ray sources and
consequently constrain the parameter space of SUSY.

Generally the ground-based detectors have the threshold energy at
the order of $100$ GeV. We assume below all the EAS detectors have such
threshold energy.
Assuming these detectors have large enough effective 
areas and have the sensitivities $I_{th}=10^{-11}, 10^{-12}, 10^{-13},
10^{-14}$ photons $cm^{-2} s^{-1}$ respectively, which are comparable
to the sensitivities of present ACTs. The sensitivity
$I_{th}$ is defined as the minimal integrated $\gamma$-ray flux
above the threshold energy that can be observed by
the detector with high significance, for example $5\sigma$,
in a finite observation time, such as $1$ to $10$ years. 

In Fig. \ref{general} we show how the SUSY parameter space is constrained
by the EAS detectors with different sensitivities. 
The points in the figure represent different SUSY models
when scanning in the SUSY parameter space. 
These models above the curves predict more than one such $\gamma$-ray 
sources should be detected at the same significance level in defining $I_{th}$.
For the subhalo model we adopt the analytic model by
Bullock et al. \cite{bullock01}. 
In the upper panel we assume the neutralino is produced
thermally while the lower panel assumes the nonthermal production. Both
the NFW and Moore profiles are shown in the same figure. We can see that
constraints assuming Moore profile are similar to the NFW case with one
order of higher sensitivity. Similarly assuming nonthermal produced also
leads to more strict constraints.

In Fig. \ref{gene} we show how the constraints depend on the
subhalo models. We assume a detector with the sensitivity of
$I_{th}=10^{12}$ photons $cm^{-2} s^{-1}$, the dark matter is
produced nonthermally and has Moore profile. We notice that for
real subhalos the constraint is more severe than these models for
distinct small halos. 

From the Figs. \ref{general} and \ref{gene} we see that the 
EAS detectors can indeed constrain the SUSY models via the observation
of $\gamma$-rays by dark matter annihilation in the subhalos due to 
the virtue of large field of view and high duty circle.
Considering that GLAST is superior only at low energies than the
EAS detectors we expect the ground EAS detectors and GLAST are
complementary to each other in the detection of dark matter annihilation
in subhalos.

\subsection{ the ARGO and HAWC experiments }

The ARGO-YBJ experiment, locates at YangBaJing
(90.522$^\circ$ east, 30.102$^\circ$ north, 4300m a.s.l)
in Tibet, China, is a ground-based telescope optimized
for the detection of small size air showers. The energy
threshold of the detector is designed to be about 100GeV.
The detector consists of a single layer of RPCs floored
in a carpet structure covering an area of $\sim 10^4 m^2$.
The detector is under construction and the central carpet has
been completed. ARGO will begin stable data taking soon after.

\begin{figure}
\includegraphics[scale=1]{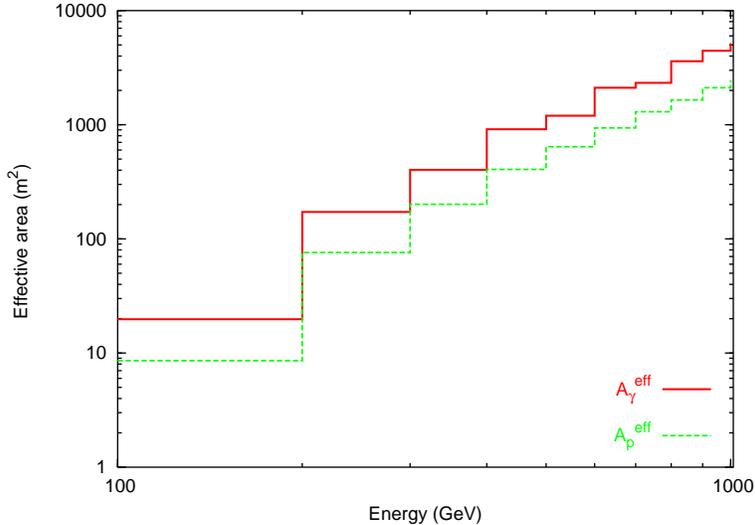}
\caption{\label{area}
The effective area of the ARGO array to primary gamma rays
and nuclei background as a function of energy. $N_{pad} \ge 20$
was adopted in simulation.
}
\end{figure}
 
The effective area of the detector characterizes the power 
of the detector in recording the number of events.
The effective area of the ARGO array is determined by 
a full Monte Carlo simulation.
We simulated $N$ showers uniformly
distributed over a large sampling area $A_s$ including the detector
and selected those which satisfy the trigger conditions.
The effective area is defined as
\begin{equation}
A_{eff}=\frac{n}{N}A_s\ ,
\end{equation}
where $N$ is the total sample events and $n$ is the number of
events satisfying the trigger conditions.
In our simulation, the software package CORSIKA \cite{corsika} is used to
simulate the shower development of the gamma 
ray signals and the nuclei background in the atmosphere and ARGOG based on
GEANT3\cite{argog} for the response of the detector to the EAS events.
To get better reconstructed events, we require the number of fired 
pad $N_{pad} \ge 20$ and the zenith angle $\theta < 45^\circ$.
A sampling area of $A_s=350 m\timesÁ350 m$, which is large enough 
for the ARGO array with $111.26 m\times 99.04 m$ under the
trigger condition  $N_{pad} \ge 20$, was used to
enclose the ARGO array at its center.
In Fig. \ref{area} we give a zenith angle averaged effective area 
as if the source was spread uniformly over the sky 
between zenith angles of 0 and 45 degrees.
We notice that above the threshold energy the effective area increases
rapidly and reaches about 5,000 $m^2$ for TeV gamma rays.
At the same time, simulation also shows that
nucleus have lower trigger efficiency than photons
leading to a suppression of the background.
%


While the Tibet group 
demonstrated the importance of a high-altitude site 
for the EAS array the 
Milagro observatory has pioneered the use of a large area water
\v{C}herenkov detector for the EAS detection and proven the
efficacy of the technique and its ability to reject the nuclei
background \cite{atkins2003}.
Combining the advantages, a next generation 
high altitude water \v{C}herenkov (HAWC) detector for VHE gamma
ray detection has been proposed recently \cite{hawc}.
The HAWC detector with the all-sky and high-duty factor capabilities,
but with a substantially lower energy threshold and a greatly increased
sensitivity would dramatically improve our knowledge of the VHE gamma
universe.

With an altitude above 
$\sim$4500 m and a large detection area of $\sim$40,000 m$^2$
the energy threshold of HAWC is as low as 50 GeV and angular resolution of 
0.25 degrees at median energy.
The average effective area of HAWC  between
zenith angles of 0 and 45 degrees for primary photons
is given in \cite{hawc} by Monte Carlo simulation.
We fit their result of the effective area as function of energy
as 
$A_{eff}=2.85\cdot 10^5 \mbox{m}^2 E^{3}/(E+195)^{3.11}$
for the energy in unit of GeV. We estimate the effective area
of HAWC to the nuclei background by assuming a same suppression
factor relative to that of primary photons as the ARGO, 
i.e., $A^{eff}_{\gamma}/A^{eff}_{p}\approx 2$.

The ability of eliminating the background further by shower shape
analysis in HAWC is simulated \cite{hawc}.
A quality factor of 1.6, which is the relative improvement in 
sensitivity of the detector, is produced independent of angular
resolution.
If combined with the angular resolution the HAWC can greatly improve the
ability of background rejection.

\subsection{ sensitivities of ARGO and HAWC }

Taking the effective area of ARGO and HAWC into the Eq. (\ref{ngamma}) and
adopting the spectrum of dark matter annihilation 
we get the signal events for the observation time 
as a function of neutralino mass.
We get the events of background similarly and the sensitivity of
ARGO and HAWC 
on looking for dark matter signals. We can then study how the SUSY models 
are constrained by these two detectors similar to the discussions given
in the subsection \ref{gdis}.

\begin{figure}
\includegraphics[scale=0.8]{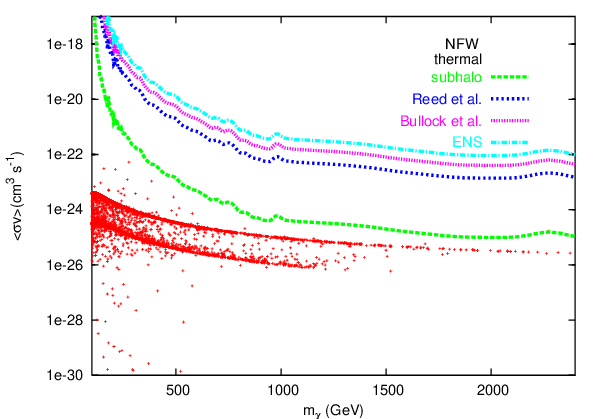}
\includegraphics[scale=0.8]{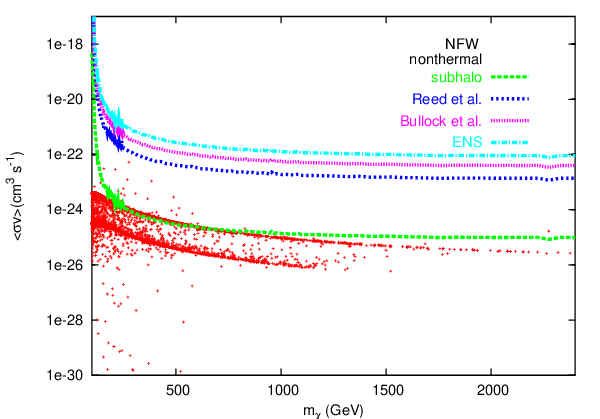}
\caption{\label{argo_nfw}
Constraints of ARGO on
the SUSY parameter space if no gamma sources are found at
the 2$\sigma$ level. The left panel gives the constraints
assuming that the dark matter are produced thermally while
the right is for nonthermal production.
NFW profile is adopted for the subhalos.
}
\end{figure}

\begin{figure}
\includegraphics[scale=0.8]{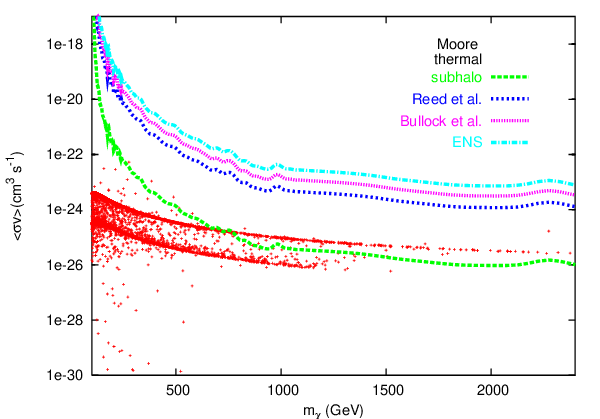}
\includegraphics[scale=0.8]{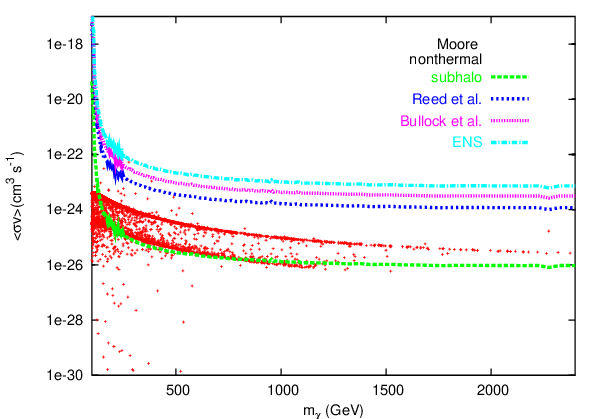}
\caption{\label{argo_m99}
Similar as Fig. \ref{argo_nfw} except that
the Moore profile is adopted for the subhalos.
}
\end{figure}
                                                                                
In Fig. \ref{argo_nfw} we give the constraints of ARGO on
the SUSY parameter space if no gamma sources are found at
the 2$\sigma$ level for 10 years observation. 
The left panel gives the constraints
assuming that the dark matter are produced thermally while
the right is for nonthermal production.
We assume a NFW profile of the subhalos. 
Fig. \ref{argo_m99} gives similar results assuming the
dark matter profile is the Moore profile.

\begin{figure}
\includegraphics[scale=0.8]{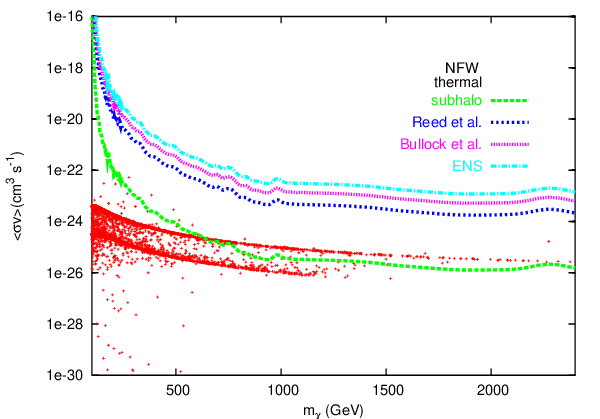}
\includegraphics[scale=0.8]{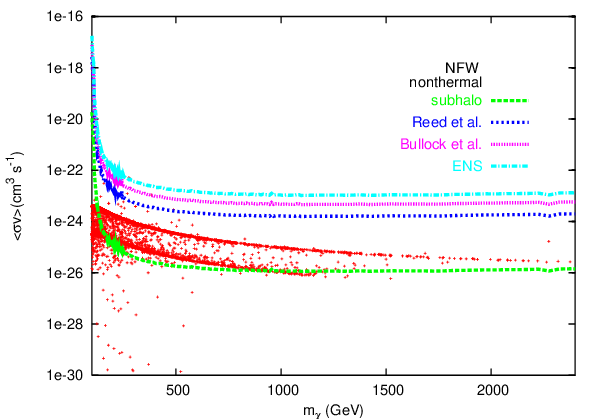}
\caption{\label{hawc_nfw}
Constraints of HAWC on
the SUSY parameter space if no gamma sources are found at
the 2$\sigma$ level. The left panel gives the constraints
assuming that the dark matter are produced thermally while
the right is for nonthermal production.
NFW profile is adopted for the subhalos.
}
\end{figure}
                                                                                
\begin{figure}
\includegraphics[scale=0.8]{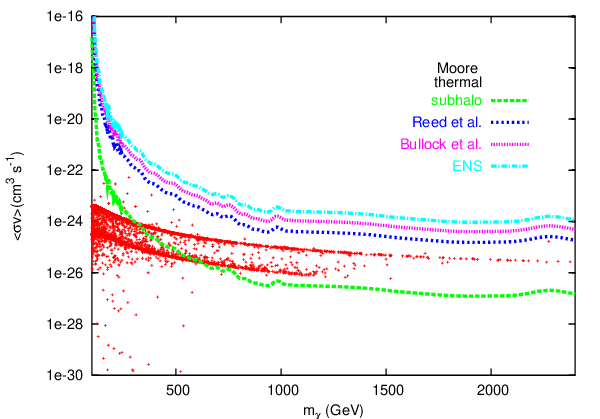}
\includegraphics[scale=0.8]{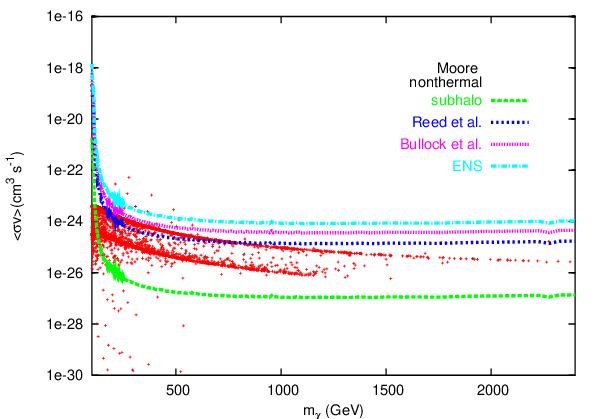}
\caption{\label{hawc_m99}
Similar as Fig. \ref{hawc_nfw} except that
the Moore profile is adopted for the subhalos.
}
\end{figure}
                                                                                
In Figs. \ref{hawc_nfw} and \ref{hawc_m99} we show the constraints
on the SUSY parameter space from HAWC for 5 years observation.
We can see that the sensitivity is greatly improved compared with ARGO.
If neutralino is produced nonthermally most SUSY parameter space will
by constrained by HAWC for the case of `subhalo' model.

\section{Summary and Conclusion}

In the work we discuss the possibility to search the dark matter
annihilation gamma rays in the subhalos of the MW. The absolute
flux from the subhalos may be smaller than that from the galactic
center, however the subhalos are also less influenced by the baryonic
matter. For heavy dark matter particles which may produce high energy
and low flux the ground based detectors with wide field of view are
complementary to the satellite based detectors, such as GLAST \cite{kou}.

Based on the N-body simulation results and scanning in the SUSY parameter
space we calculate the flux of gamma rays from the subhalos. We then
discuss the possibility of detection of these fluxed at ground EAS
detectors, especially the ARGO and HAWC experiments.
The properties of ARGO and HAWC are studied by Monte Carlo simulations.
Due to our results the SUSY parameter space is constrained
if no gamma signal from the subhalos is detected.
In conclusion, the ground based detectors have the capability to
search the dark matter signal and constrain the SUSY parameter space,
complementary to the GLAST.

\begin{acknowledgments}
This work is supported by the NSF of China under the grant
No. 10575111 and supported in part by the Chinese Academy of
Sciences under the grant No. KJCX3-SYW-N2.
\end{acknowledgments}

\end{document}